\documentstyle[pra,aps]{revtex}
\begin{document}
%
%
\title{Static Properties of the Trapped Bose-Fermi 
       Mixed Condensate of Alkali Atoms}
\author{T. Miyakawa, K. Oda, T. Suzuki and H. Yabu}
\address{Department of Physics, Tokyo Metropolitan University, 
         Hachioji, Tokyo 192-0397, Japan}
\date{\today}
\maketitle
\begin{abstract}
Static properties of a bose-fermi mixture 
of trapped potassium atoms are studied 
in terms of coupled Gross-Pitaevskii 
and Thomas-Fermi equations 
for both repulsive and attractive bose-fermi interatomic potentials. 
Qualitative estimates are given 
for solutions of the coupled equations, 
and the parameter regions are obtained analytically  
for the boson-density profile change 
and for the boson/fermion phase separation. 
Especially, the parameter ratio $R_{int}$ is found 
that discriminates the region of the large boson-profile change. 
These estimates are applied for numerical results 
for the potassium atoms and checked their consistency.   
It is suggested that a small fraction of fermions could be trapped 
without an external potential for the system 
with an attractive boson-fermion interaction.
\end{abstract}
\pacs{PACS number: 03.75.Fi, 05.30.Fk,67.60.-g}
%
%
\section{Introduction}

Recent development in the cooling and trapping technique of atoms 
and the following success 
in achieving the Bose-Einstein condensation 
of alkali atoms\cite{bec,review,dalfovo} 
have now opened up a new era in the 
progress in quantum physics. 
While the condensed system serves as a testing ground 
for a research in fundamental problems of quantum mechanics, 
it also offers a new example of finite quantum many-body systems 
such as hadrons, nuclei and microclusters. 
An important characteristic of the trapped alkali atoms is 
that it is a dilute system of weakly interacting particles 
and an ideal place to test genuine properties 
of condensed systems predicted by theories. 

Along with a further progress 
in the study of Bose-Einstein condensed systems, 
a similar technique is being extended 
to create a degenerate gas of fermionic atoms, 
where a number of theoretical studies have been made\cite{fermi}. 
To realize such a degenerate fermionic system
that requires a still lower temperature than bose systems, 
the technique of sympathetic cooling 
has been investigated\cite{sympa}: 
the cooling mechanism through the collisions 
with coexisting cold bose particles 
in a polarized boson-fermion mixture,  
where the fermion-fermion interaction is less effective. 

The mixture of bose and fermi particles is itself 
an interesting system for a study:  
the hydrogen-deuterium system has been studied 
already at the early stage of these investigations\cite{hydro}.
Recently, theoretical studies have been made 
for the trapped boson-fermion system 
of alkali atoms 
and proposed many interesting properties: 
the exotic density configurations 
through the repulsive or attractive 
boson-fermion interactions \cite{molmer,ng}, 
the decaying processes 
after the removal of the confining trap \cite{stringari}, and
the metastability of the ${}^7$Li-${}^6$Li mixture \cite{minniti}.  

In this paper, 
we study static properties of a mixed bose-fermi system 
of trapped potassium atoms, 
where one fermionic (${}^{40}$K) 
and two bosonic isotopes (${}^{39}$K and ${}^{41}$K) are known 
as candidates for a realization of such a degenerate quantum system. 
Trapping of the fermionic isotope ${}^{40}$K 
has already been reported\cite{kfe}. 
These isotopes are also of an interest 
since the boson-fermion interaction may be repulsive 
or attractive depending on their choice\cite{pota,cote}. 

Below we first describe a set of equations 
for the mixed system at $T=0$, 
where the bosonic part is given by a Gross-Pitaevskii (GP) equation 
and including boson-fermion interaction, 
and the fermionic part is by the Thomas-Fermi (TF) equation. 
The use of the GP equation allows one to study a system 
with a net attractive potential for bosons 
where the Thomas-Fermi-type (TF) approximation is not applicable. 
The applicability of these equations for the trapped potassium system 
was discussed in \cite{ng}.  

In sec. III, we discuss qualitative properties 
of solutions from the GP and TF equations analytically;
especially, their dependence on the boson/fermion number 
is estimated. 

In ref. \cite{ng}, 
general features of solution including a deformed system
is discussed in the repulsive boson-fermion interaction.
On the other hand, to study more realistic system, 
we concentrate on the specific systems of potassium isotopes, 
i.e.,${}^{39}$K-${}^{40}$K system and ${}^{41}$K-${}^{40}$K system
using empirically derived atomic interactions.  
We also consider the effect 
of changing parameters of the trapped potential. 
Finally, we mention the possibility 
of trapping fermions 
only through the attractive interactions 
by the trapped bose particles. 
%
%
\section{Formulation}
We consider a spin-polarized system of bosons and fermions at $T=0$ 
described by a Hamiltonian $H =H_b +H_f +V_{bf}$: 
\begin{mathletters}
\begin{eqnarray}
   H_b &=& \int\! d\vec{r} 
           \phi^\dagger(\vec{r}) 
             (-\frac{\hbar^2}{2 m_b} \nabla^2
              -\mu_b +\frac{1}{2} m_b \omega_b^2 r^2) 
           \phi(\vec{r}) 
\nonumber\\
       & &+\frac{1}{2} g \int\!\int\! d\vec{r} d{\vec{r}\,'} 
           \phi^\dagger(\vec{r}) \phi^\dagger({\vec{r}\,'}) 
             \delta(\vec{r} -{\vec{r}\,'}) 
           \phi({\vec{r}\,'}) \phi(\vec{r}), 
\label{eQhA}\\
   H_f &=& \int\! d\vec{r} 
           \psi^\dagger(\vec{r}) 
             (-\frac{\hbar^2}{2 m_f} \nabla^2 -\mu_f 
              +\frac{1}{2} m_f \omega_f^2 r^2) 
           \psi(\vec{r}), 
\label{eQhB}\\
   V_{bf} &=& h \int\!\! d\vec{r} d{\vec{r}\,'} 
              \phi^\dagger(\vec{r}) \psi^\dagger({\vec{r}\,'})
                \delta^3(\vec{r} -{\vec{r}\,'}) 
              \psi({\vec{r}\,'}) \phi(\vec{r}), 
\label{eQhC}
\end{eqnarray}
\end{mathletters}
where $\phi(\vec{r})$ and $\psi(\vec{r})$ respectively denote boson 
and fermion field operators with masses $m_b$ and $m_f$. 
Because of diluteness, 
the particle interactions have been approximated
by the $s$-wave-dominated contact potential
$v(\vec{r} -{\vec{r}\,'}) 
=\{g {\rm\ or\ } h\} \delta^3(\vec{r} -{\vec{r}\,'})$.   
The strengths of the coupling constants $g$ (boson-boson) 
and $h$ (boson-fermion) are given by 
\begin{equation}
    g ={4\pi \hbar^2 \over m_b} a_{bb},     \qquad
    h ={2\pi \hbar^2 \over m_{bf}} a_{bf},  \label{eQe}
\end{equation}
where $a_{bb}$ and $a_{bf}$ are the $s$-wave scattering lengths
for the boson-boson and boson-fermion scatterings
and $m_{bf} =m_b m_f/(m_b +m_f)$ is a reduced mass. 
The fermion-fermion interaction has been neglected 
because of diluteness and particle polarization. 
The chemical potentials $\mu_b$ and $\mu_f$ 
in (\ref{eQhA}) and (\ref{eQhB}) are determined 
from the condensed boson/fermion numbers $N_b$ and $N_f$ 
through the ground state expectation values:
\begin{eqnarray}
   N_b ={\langle \int\!\! d\vec{r}\, 
                 \phi^\dagger(\vec{r}) \phi(\vec{r})
         \rangle},                                   \qquad
   N_f ={\langle \int\!\! d\vec{r}\, 
                 \psi^\dagger(\vec{r}) \psi(\vec{r})
         \rangle}. 
\label{eQh}
\end{eqnarray}

In the mean-field approximation at $T=0$, 
the bosons occupy the lowest single-particle state $\varphi_b(\vec{r})$, 
and the energy of the system is just given 
by a replacement of field operator $\phi(\vec{r})$ 
in the Hamiltonian with its expectation value: 
\begin{equation}
   \Phi(\vec{r}) \equiv {\langle \phi(\vec{r}) \rangle} 
                  =\sqrt{N_b} \varphi_b(\vec{r})
\end{equation}
that describes the order parameter of the Bose-Einstein condensate. 
(We here neglect quantities of the order $O(N_b^{-1})$.) 
In this approximation, 
the boson density $n_b(\vec{r})$ is given by $|\Phi(\vec{r})|^2$. 

In the same approximation, 
the fermion wave function is given 
by a Slater determinant;  
the single-particle states are determined, e.g., 
by the Hartree-Fock self-consistent equation. 
In the present approximation 
of neglecting fermion-fermion interactions,
we only have to solve for the single particle states 
under the effective potential for fermions: 
the trapping potential and the boson-fermion interaction. 
In the actual calculation, we however adopt a 
semiclassical (TF) description 
for fermion density,  
which is known to provide a good approximation 
as far as the number of atoms is sufficiently large\cite{fermi}. 

We thus obtain a set of equations for $\Phi$ and $n_f$ \cite{ng}: 
\begin{eqnarray}
      && [-\frac{\hbar^2}{2 m_b} \nabla^2 
          +\frac{1}{2} m_b \omega_b^2 r^2
          +g\, n_b(\vec{r}) 
          +h\, n_f(\vec{r})] \Phi(\vec{r}) 
          =\mu_b \Phi(\vec{r}), 
\label{GPeq}\\
      && \frac{\hbar^2}{2 m_f} 
         [6 \pi^2 n_f(\vec{r})]^{2/3} 
          +\frac{1}{2} m_f \omega_f^2 r^2
          +h\, n_b(\vec{r}) =\mu_f, 
\label{Fermieq}
\end{eqnarray} 
where the boson and fermion densities are defined by 
$n_b(\vec{r}) ={\langle \phi^\dagger(\vec{r}) 
                        \phi(\vec{r}) \rangle}
              \sim |\Phi(\vec{r})|^2$ 
and 
$n_f(\vec{r}) ={\langle \psi^\dagger(\vec{r}) 
                        \psi(\vec{r}) \rangle}$.  
%
%
\section{Qualitative Study of Solutions}
     Before the study of numerical results 
for (\ref{GPeq}) and (\ref{Fermieq}), 
we discuss about their solutions qualitatively.    

\vspace{0.5cm}
\noindent
i) {\it Radius of the boson/fermion distribution}  
\smallskip

If we neglect the boson-fermion interaction, 
we would obtain a much broader distribution for fermions 
as compared with that for bosons 
because of the exclusion principle. 
To see this, 
we consider a free boson/fermion system
trapped in a harmonic oscillator potential.  
The mean square radius of each system is obtained 
from the virial theorem:
\begin{equation}
   {\langle r^2 \rangle}_b /N_b =\frac{3}{2} \xi_b^2,         \qquad
   {\langle r^2 \rangle}_f /N_f =\frac{3}{4} (n_F+2) \xi_f^2, 
\label{eQd}
\end{equation}
where the harmonic-oscillator lengths $\xi_{b,f}$ are 
\begin{equation}
     \xi_{b,f} \equiv \sqrt{\frac{\hbar}{m_{b,f} \omega_{b,f}}}.  
\label{eQc}
\end{equation}
and $N_{b,f}$ are the total boson/fermion numbers.  
The $n_F$ in (\ref{eQd}) is the number of fermions 
in the last occupied shell 
that is obtained as a solution of  
$N_f =(n_F +1) (n_F +2) (n_F +3) /6 \sim n_F^3$. 
The ratio of the root-mean-square (rms) radii 
$\sqrt{{\langle r^2 \rangle}_f/{\langle r^2 \rangle}_b}$ 
thus increases 
as the power $N^{1/6}$ when 
$N =N_b \sim N_f$ and $\xi_b \sim \xi_f$. 
In fact, for a repulsive boson-boson interaction, 
the boson distribution become broader than 
that for a free system just discussed, 
but this effect does not change the qualitative structure 
as shown in sec. V. 
On the other hand, 
the strong boson-fermion interaction may give 
qualitative changes of the density profiles 
such as the phase separation and the fermion collapse 
as shown in \cite{molmer,ng}.  
We also discussed about them 
from different points of view in iii) in this section. 

\vspace{0.5cm}
\noindent
ii){\it Scaled GP and TF equations and 
changes of the density distribution for the different $N_{b,f}$}
\smallskip

For qualitative estimations 
of the density distribution, 
we should use the scaled dimensionless variables \cite{dalfovo}:
\begin{equation}
     \vec{x}           =\frac{\vec{r}}{\xi_b},            \quad
     \rho_{b,f}        =\frac{\xi_b^3}{N_{b,f}} n_{b,f},  \quad
     \tilde{\mu}_{b,f} =\frac{2}{\hbar\omega_b} \mu_{b,f}, 
\label{eQb}
\end{equation}
where $N_{b,f}$, $\xi_{b,f}$ are the total boson/fermion numbers 
and the harmonic-oscillator lengths 
defined in eqs. (\ref{eQh}) and (\ref{eQc}).  
Using these scaled variables, 
the GP and TF equations in eqs. (\ref{GPeq}) and (\ref{Fermieq}) 
become
\begin{eqnarray}
   && [-\nabla_{x}^2+ x^2
       +\tilde{g}\, N_{b}\, \rho_b(\vec{x})
       +\tilde{h}\, N_{f}\, \rho_f(\vec{x})]
      \tilde{\varphi_b}(\vec{x}) 
       =\tilde{\mu}_b \tilde{\varphi_b}(\vec{x})  
\label{scGPeq}\\
   && \frac{1}{R_m} 
      [6 \pi^2 N_{f} \rho_f(\vec{x})]^{\frac{2}{3}}
      +R_m R_\omega^2 x^2
      +\tilde{h}\, N_b\, \rho_b(\vec{x}) 
      =\tilde{\mu}_f,           
\label{scFermieq}
\end{eqnarray} 
where $R_m =m_f/m_b$ and $R_\omega =\omega_f/\omega_b$. 
The scaled coupling constants $\tilde{g}$ and $\tilde{h}$ 
in eqs. (\ref{scGPeq}) and (\ref{scFermieq}) 
are defined by 
\begin{equation}
     \tilde{g} \equiv \frac{2g}{\hbar \omega_b \xi_b^3} 
               =8\pi \frac{a_{bb}}{\xi_b},              \quad
     \tilde{h} \equiv {2h \over \hbar \omega_b \xi_b^3}
               =4\pi \frac{a_{bf}}{\xi_b} 
                \left( 1 +\frac{1}{R_m} \right), 
\end{equation}
where eq. (\ref{eQe}) has been used 
to obtain the $a_{bb,bf}$-representations, 
and the last factor comes from 
$m_b/m_{bf} =1 +R_m^{-1}$.  

From eqs. (\ref{scGPeq}) and (\ref{scFermieq}), 
we should notice that the scaled equations include 
six independent parameters, 
($\tilde{g} N_b$, $\tilde{h} N_f$, $N_f$, $\tilde{h} N_b$,
$R_m$, $R_\omega$); 
for two systems where these parameters take the same values,  
their physical properties become similar under the scaling relations 
(\ref{eQb}).    

Here, based on eqs. (\ref{scGPeq}) and (\ref{scFermieq}), 
we discuss about changes of the density distribution 
by the scaled parameters. 
In this subsection, we assume $R_m =R_\omega =1$ 
and the large particle numbers $N_b \sim N_f \sim N \gg 1$; 
for the effect of $R_\omega \neq 1$, 
the discussion will be given in sec. V.   

Let us consider the fermion TF equation (\ref{scFermieq}), first. 
For the chemical potential $\tilde{\mu}_f$, 
we use the result in the system 
without boson-fermion interaction: 
$\tilde{\mu}_f \sim 2(6N_{F})^{1/3} \sim N^{1/3}$. 
It should be a good approximation when $N_f \gg 1$. 
Using it in eq. (\ref{scGPeq}), 
we obtain 
$\rho_f \sim N_f^{-1/2}$ 
at the central region ($x \sim 0$). 
On the other hand, 
the boson equation (\ref{scGPeq}) gives 
$\rho_b \sim (\tilde{g} N_b)^{-3/5}$ 
under the TF approximation valid in $N_b \gg 1$.  
Using these estimates, 
we can obtain the ratio $R_{int}$ of 
the boson-boson/boson-fermion interaction effects 
in the GP equation (\ref{scGPeq}): 
\begin{equation}
     R_{int} \equiv \left. 
                    \frac{\tilde{h} N_f \rho_f}{
                          \tilde{g} N_b \rho_b} 
                    \right|_{x=0} 
             \sim \frac{\tilde{h} N_f}{\tilde{g} N_b}
                  \frac{N_f^{-\frac{1}{2}}}{
                        (\tilde{g} N_b)^{-\frac{3}{5}}} 
             \sim \frac{\tilde{h} N_f}{\tilde{g} N_b}
                  \left(\frac{\tilde{g} N_b}{N_f} 
                  \right)^{\frac{1}{2}}. 
\label{eQi}
\end{equation}
The small $R_{int}$ ($\ll 1$) indicates the small contribution 
from the boson-fermion interaction for the boson part, 
and we obtain $\rho_b \sim \rho_b(h=0)$. 
However, when $R_{int} \geq 1$, 
the boson-fermion interaction is not negligible, 
and the boson density profile can be quite different 
from that in non-interacting case ($h =0$).

In ${}^{39}$K-${}^{40}$K system discussed in sec. V, 
the scaled parameters are 
\begin{equation}
     (\tilde{g} N_b, \tilde{h} N_f, N_f, \tilde{h} N_b) 
          =(2.68 \times 10^{-2} N_b, 
            1.57 \times 10^{-2} N_f,
            N_f, 
            1.57 \times 10^{-2} N_b).   
\label{eQj}
\end{equation}
In $N \sim N_b \sim N_f$, 
$\tilde{g} N_b$ and $\tilde{h} N_f$ are in the same order,  
but $N_f$ is a factor of 2 larger than $\tilde{h} N_f$, 
so eq. (\ref{eQi}) gives 
$R_{int} \sim 10^{-1}$. 
It explains the rather small change of the boson density profile 
$\rho_b \sim \rho_b(h=0)$ in ${}^{39}$K-${}^{40}$K system, 
as shown in the numerical results in sec. V. 

It is very interesting to apply the above estimation 
for the results by \cite{molmer,ng} 
and compare them with the results in this paper.    
One of the parameter set with which the boson density profiles 
changed drastically is 
\begin{equation}
     (\tilde{g} N_b, \tilde{h} N_f, N_f, \tilde{h} N_b) 
          =(4.22 \times 10^6, 
            4.22 \times 10^6 {h /g},
            10^6, 
            4.22 \times 10^6 {h /g}),  
\label{eQf}
\end{equation}
where $h/g =\tilde{h}/\tilde{g} =0 \sim 5/4$ has been taken 
and the qualitative changes of the boson density profile 
have been seen $h/g \geq 3/4$ \cite{molmer,ng}. 
It should be noticed that the parameters in (\ref{eQf}) 
are in the same order when $h/g \sim 1$, 
and it means $R_{int} \sim 1$ in that case. 
For this reason, the boson density profile changed 
drastically from that of $h=0$ with these parameters. 

To check the relation between the boson profile change 
and the ratio $R_{int}$ in eq. (\ref{eQi}) a little more, 
we solved eqs. (\ref{scGPeq}) and (\ref{scFermieq}) 
numerically for a) $N_f =10^6$ and b) $N_f =10^8$ 
with other parameters fixed as in (\ref{eQf}), 
and, for $h/g$, two values were taken $h/g =0, 0.6$. 
The resultant boson densities are shown in fig.~1: 
$n_b(h \neq 0)$ has very different profile from $n_b(h=0)$ in case a), 
but the very small change is found between them in case b). 
Those results shows the clear relation between $R_{int}$ and 
the boson density profile. 

In summary, 
the parameter regions exist where the boson density profile 
changes very large or not, 
and these regions are discriminated by the value of $R_{int}$. 

We should comment that the results in this paper 
are complementary with those with (\ref{eQf}) in \cite{molmer} 
in the scale of the ratio $R_{int}$.

\vspace{0.5cm}
\noindent
iii) {\it Boson-fermion interaction effects 
for the fermion distribution function}
\smallskip

We next consider the effect of the boson-fermion interaction 
on the fermion distribution, 
especially the phase separation phenomena 
in the strongly repulsive boson-fermion interaction case, 
which has been proposed in \cite{molmer,stringari}. 
Here we assume the TF approximation for bosons 
(neglecting the kinetic term in eq.(\ref{Fermieq})). 
As discussed in the literature \cite{molmer,ng,stringari}, 
the effective potential for fermions receives 
an additional repulsive or attractive contribution from bosons 
within the range 
$r \leq R_b \equiv \sqrt{2 \mu_b /m_b \omega_b^2}$ 
depending on the sign of the interaction. 

For a strongly repulsive boson-fermion interaction ($h>0$), 
the fermion will be squeezed out from the center, 
so that the two kind of particles tend 
to make a separate phase \cite{molmer,ng}. 
Here we give analytical estimation for this interesting phenomena. 
As shown in \cite{molmer,ng}, in the phase-separated system,    
the fermion are almost completely pushed away 
outside the boson distribution, 
so that the critical ratio $(h/g)_c$ for the phase separation 
can be estimated from the vanishing fermion density 
at the center $n_f(r=0) =0$. 
Neglecting the boson-fermion interaction 
in the boson TF equation, 
we obtain  
\begin{equation}
    N_f = F(\alpha) 
               \equiv 
             \frac{1}{3\pi} 
             \left( {\mu_{b} \over \hbar \omega_{b}} \right)^3
             \left\{ \alpha (\alpha -2) (\alpha -1)^{1/2} 
               +\alpha^3 \arctan(\alpha-1)^{1/2} \right\}  
\label{eQa}
\end{equation}
where $\alpha =h /g \geq 1$.
In fig.~2, the function $F(\alpha)$ in (\ref{eQa}) is plotted 
when $N_f =10^3$ with the parameters used in \cite{ng}: 
($\tilde{g}N_{b}$, $\tilde{h}N_{f}$, 
 $N_{f}$, $\tilde{h}N_{b}$, 
 $R_{m}$, $R_{\omega}$)
=($3.0\times 10^{4}$, $3.0\times 10 \alpha$, 
  $10^{3}$, $3.0\times 10 \alpha$, $1$, $1$) \cite{FNA}. 
From the crossing point of $f(\alpha)$ and $N_f =10^3$ (dashed line) 
in fig.~2, 
the critical point can be read off: 
$\alpha =(h/g)|_{c} \sim 1.3$, 
which is almost consistent 
with the result given in \cite{ng} ($\alpha \simeq 1.18$). 

For an attractive interaction ($h<0$), 
fermions tend to concentrate 
and increase the overlap with bosons. 
Thus we may expect a coherence of the two kinds of particles 
to occur in this case. 
For large attraction, another interesting phenomena, 
fermion collapse in mixture has been proposed \cite{molmer}.
%
%
\section{Numerical Procedure}
%
%
In the numerical calculation, 
we solved the set of equations (\ref{GPeq}) and (\ref{Fermieq}) 
by reducing them into the following equation 
for the order parameter:
\begin{equation}
 \left[ t \frac{d^2}{dt^2} 
       +\left( \frac{3}{2} -t \right) \frac{d}{dt}
       -\frac{1}{4} \left\{ 3 -\tilde{\mu}_b  
                              +\tilde{g} \frac{f(t)^2}{e^t} 
                    \right\}
       -\frac{\tilde{h}}{24 \pi^2} 
        \left\{ \tilde{\mu}_f 
               -\frac{m_f \omega_f^2}{m_b \omega_b^2} t 
               -\tilde{h} \frac{f(t)^2}{e^t} 
        \right\}^{3/2} 
 \right] f(t)=0, 
\label{numeric}
\end{equation}
where 
$t =x^2 =(r/\xi_b)^2$ 
and the scaled parameters 
$x$, $\tilde{g}$, $\tilde{h}$ and $\tilde{\mu}_{b,f}$
have been defined in (\ref{eQb}). 
The function $f$ is related to the order parameter 
by $\Phi(r) / r=e^{-x^2} f(x^2)$. 
The boundary conditions for $f(t)$ are given 
by the value at $t=0$ 
and by the asymptotic condition 
$f(t \rightarrow \infty) \sim t^{(\tilde{\mu}_b -3) /4}$.  
We first give initial values for $f(0)$ and $\mu_f$ 
and solved the equation (\ref{numeric}) numerically 
by means of the relaxation method. 
Finally, the fermion density $n_f(r)$ is obtained 
from eq.(\ref{Fermieq}). 
The details of the calculation will be given elsewhere\cite{miya}. 

We consider the potassium atoms 
where there exist two bosonic (${}^{39}$K, ${}^{41}$K) 
and one fermionic (${}^{40}$K) isotopes. 
The precise values of the interatomic 
interaction are not known, 
but the recent estimate from molecular scattering\cite{pota,cote} 
suggest a repulsive interaction between ${}^{39}$K and ${}^{40}$K, 
and an attractive one between ${}^{41}$K and ${}^{40}$K, 
although the values still have rather large ambiguities. 
We take up the following values 
for the $s$-wave scattering lengths \cite{cote}: 
$a_{bb}(\hbox{${}^{39}$K-${}^{39}$K}) = 4.29{\rm\,nm}$, 
$a_{bf}(\hbox{${}^{39}$K-${}^{40}$K}) = 2.51{\rm\,nm}$, 
$a_{bb}(\hbox{${}^{41}$K-${}^{41}$K}) =15.13{\rm\,nm}$, 
$a_{bf}(\hbox{${}^{41}$K-${}^{40}$K}) =-8.57{\rm\,nm}$. 
We take the atomic masses to be the same for all the isotopes: 
$m_b =m_f =0.649\times 10^{-25} {\rm\,kg}$.
The angular frequency $\omega_b$ of the bosonic external potential 
is fixed at $100 {\rm\,Hz}$, 
while $\omega_f$ is allowed to vary. 
In this case, it is corresponding to 
$\tilde{g} =2.68\times 10^{-2}$,
$\tilde{h} =1.57\times 10^{-2}$ for ${}^{39}$K-${}^{40}$K system and
$\tilde{g} =9.44\times 10^{-2}$,
$\tilde{h} =-5.34\times 10^{-2}$ for ${}^{41}$K-${}^{40}$K system.
%
%
\section{Results}
%
%
First, we consider the ${}^{39}$K-${}^{40}$K system 
where the boson-fermion interaction is repulsive. 
In this case, because of $\tilde{h} N_b/\tilde{g} N_{b} <1$, 
phase separation do not occur.
Fig.~3 shows the density-distribution functions
of bosons (a) and fermions (b) 
for $N_b =10000$, $N_f =1000$ 
and $\omega_f =\omega_b =100{\rm\,Hz}$. 
The dashed line shows the result for $a_{bf}=0$ ($h=0$)
as compared with the solid line for 
$a_{bf} \neq 0$ ($h \neq 0$). 
As seen from the figure, 
the fermions have a much broader distribution 
than bosons even for a much less number of particles.  
That has been discussed in sec. III-i). 
The boson-fermion interaction is seen to squeeze out the fermions 
leading to a fermion-density depletion at the center, 
while its effect on the boson distribution is negligible. 
This also can be seen in Table I 
where some observables are listed 
for the two cases 
$(N_b,N_f)$=(1000,1000) and (10000,1000) 
with/without the boson-fermion interaction. 
The robust structure of the boson distribution just mentioned 
is reflected in the almost constant values of the rms radius 
and other observables for the bosonic part 
of the mixed condensate 
against the switching on/off of the boson-fermion interaction. 
To understand this robustness, 
we apply the estimation in sec. III-ii): 
the (1000,1000) system just corresponds to the parameters 
of eq. (\ref{eQj}), 
so the $R_{int}$ in (\ref{eQi}) becomes 
$R_{int} \sim 10^{-1}$. 
For (10000,1000) case, 
the scaled parameters are 
($\tilde{g}N_{b}$, $\tilde{h}N_{f}$, 
 $N_{f}$, $\tilde{h}N_{b}$)
=($2.68\times 10^2$, $1.57\times 10$, 
  $10^3$, $1.57\times 10^2$), 
and the ratio $\tilde{h} N_f/\tilde{g} N_b$ in eq. (\ref{eQi}) 
gives additional factor $10^{-1}$, 
so we obtain 
$R_{int} \sim 10^{-2}$. 
It shows that (10000,1000) system is more robust 
than (1000,1000) case.   
 
It is supported in ref. \cite{ng} 
that the TF approximation yields qualitative correct result. 
Using the method in \cite{shuck}, 
we can also verify that the correction for the TF approximation 
is small for 1000 fermion case.

We next turn to the ${}^{41}$K-${}^{40}$K system 
where the boson-fermion interaction is attractive. 
Because of the large value of the strength, 
one may expect a sizable effect of the boson-fermion interaction 
for this system. 
Moreover, the strong boson-boson interaction makes 
a stronger overlap of bosons and fermions 
and thus the attractive interaction are more efficient.  
So the system might be expected to be unstable against collapse. 
if the effect of the boson-fermion interaction is too strong. 
Actually, 
$(N_b,N_f)$=(1000,1000) and (10000,1000), 
we obtained a distribution 
as given in fig.~4 and Table II. 
The effect of the attraction on the fermions 
is larger than the ${}^{39}$K-${}^{40}$K system, 
although it is not strong enough to cause an appreciable change 
in the boson distribution. 
The stability condition for the fermion collapse  
can be derived from the scaling law in refs. \cite{molmer}: 
\begin{equation}
     \frac{\tilde{h}^2}{\tilde{g}} \lesssim c 
                                   \equiv 2 \times 6.9 N^{-1/6}. 
\label{eQm}
\end{equation}
In the present case, we obtain 
$\tilde{h}^2/\tilde{g} =3.02 \times 10^{-2}$ 
and $c \simeq 4.4$, 
so that the stability condition (\ref{eQm}) 
is well satisfied, 
and we have found the parameters of ${}^{41}$K-${}^{40}$K system
is in (quasi) stable region.

Is the bosonic distribution always robust 
against a mixture of fermions?  
Aside from the possibility 
of changing the interaction strength 
via Feshbach resonances\cite{feshbach}, 
we have a possibility 
of enhancing the boson-fermion interaction effect 
using the different confining potential
for bosons and fermions separately (i.e. $\omega_f/\omega_b \ne 1$). 
Fig.~5 shows the boson/fermion distribution 
for (a) ${}^{39}$K-${}^{40}$K ($R_\omega =1, 5, 10$) 
and (b) ${}^{41}$K-${}^{40}$K ($R_\omega =1, 4, 7$): 
$(N_b,N_f)=(1000,1000)$, $\omega_b =100{\rm\,Hz}$. 
For a large value of the ratio $R_\omega=\omega_f/\omega_b$, 
the fermions are trapped in a strong confining potential 
and thus have a large overlap with bosons; 
the boson distribution (solid lines) is sensitive 
to the value of $R_\omega$
even though the bosonic parameters are fixed. 
This may be traced back to a rather large fraction 
of the boson-fermion interaction energy in the total boson energies: 
for the ${}^{41}$K-${}^{40}$K system at $R_\omega =7$, for instance,
the ratio of the boson-fermion interaction energy 
to the total boson energy becomes 
$E_{BF}/E_{tot}({}^{41}{\rm K}) =0.36$ \cite{FNA}, 
while it is less than 1 per cent to the total fermion energy.  
The change of the boson-density distribution can be estimated 
from the ratio $R_{int}$ in (\ref{eQi}). 
When $R_\omega \neq 1$, it becomes 
\begin{equation}
     R_{int} =\frac{\tilde{h} N_f}{\tilde{g} N_g}
              \left(\frac{\tilde{g} N_b}{N_f} 
              \right)^{\frac{1}{2}} R_\omega^{3/2}.
\end{equation}
so that the boson-density distribution become more sensitive 
for large values of $R_\omega$.  

The effect of the boson-fermion interaction on fermions may show up 
in the opposite extreme, i.e., 
for vanishing $R_\omega$. 
In this case, the fermions can keep themselves from escape only 
through the attractive boson-fermion interaction. 
This possibility can be estimated from the chemical potential: 
when fermions are trapped by bosons, 
the $\mu_f$ should take the negative value. 
Taking the TF approximation for the bosonic part 
($\tilde{g} N_b \gg 1$, $\tilde{g} N_b \gg \tilde{h} N_f$)
and setting $\omega_f=0$ for the fermionic part, 
we obtain the total fermion number represented by $\mu_f$:
\begin{equation}
     N_f =\frac{1}{48 |h/g|^{\frac{3}{2}}}
          \left\{ \tilde{\mu}_f +\left|\frac{h}{g}\right| 
                         \left( \frac{15}{8\pi} 
                                \tilde{g} N_b \right)^{2/5} 
               \right\}^3. 
\label{eQk}
\end{equation}
When $\mu_f =0$ in (\ref{eQk}), 
we obtain the maximum number for the boson-trapped fermions:  
\begin{equation}
      N_f^{max} = \frac{1}{48} 
      \left|\frac{h}{g}\right|^{3/2} 
      \left( \frac{15}{8\pi} \tilde{g} N_b \right)^{6/5} 
      = 2.8 \times 10^{-4} N_b^{6/5}
\label{eQl}
\end{equation}
where we used the parameters of ${}^{41}$K-${}^{40}$K system 
to obtain the last term. 
From eq. (\ref{eQl}), 
estimations can be obtained for the boson-trapped fermion number:
for $N_b =10000$, 
only 18 fermions are bound, 
while for $N_b =10^6$ more than 4000 fermions will be trapped 
within the bose condensate. 
This qualitative estimate is verified also
in the numerical calculation for $N_b =10000$, 
although one should actually take the shell effect into account 
for such a small number of fermions\cite{miya}. 

%
\section{Summary}
%
%
In the present paper, 
we studied static properties of a mixed system 
of bosons (${}^{39}$K or ${}^{41}$K) and fermions (${}^{40}$K) 
in the trapping potential 
at temperature zero. 
We solved a coupled GP and TF equations 
for a different combination of boson and fermion particle numbers 
and also for a repulsive or an attractive interaction 
of bosons and fermions 
depending on the combination of potassium isotopes. 

The analytical estimations have been done generally 
for the rms radii of the boson/fermion distribution, 
the parameter-dependence of their profile 
and the phase separation, 
and we have found the feasible conditions 
for them. 
These analytical estimations have been applied 
to the numerical results for the potassium system  
and also compared with the discussion done in \cite{molmer,ng}: 
consequently, the consistency was checked for these estimations. 

For similar particle numbers and external potential parameters, 
the fermions have a much more extended distribution 
and larger energies than bosons 
because of the exclusion principle. 
We studied the effect of the boson-fermion interaction 
on the density distribution of both kinds of particles. 
For a repulsive interaction, 
the fermions tend to be pushed out, 
but the effect is not very large for the 
parameters of ${}^{41}$K-${}^{40}$K system 
adopted in the present paper. 
In the case of the attractive interaction, 
a coherence of  ${}^{41}$K and ${}^{40}$K occurs 
and is strongly enhanced 
if the trapping potential is adjusted so as to make them overlap. 
It was shown also that the fermions could be kept trapped 
without external potential 
due only to an attractive interaction with trapped bose particles. 
One may as well consider the opposite case 
where the bosons are trapped only via 
an attractive interaction 
with surrounding trapped fermi particles. 
This will be discussed in a future publication\cite{miya}. 
%
%

\newpage
%
%
\begin{table}
\caption{
Results for the  ${}^{39}$K-${}^{40}$K system 
with ($a_{bf}=2.51{\rm\,nm}$) and without ($a_{bf}=0$) 
the boson-fermion interaction for $(N_b,N_f)=(1000,1000)$ 
and $(10000,1000)$. 
Contributions to the equilibrium energy 
(measured in the unit of $\hbar\omega_b/2$) per particle, 
the central density $n(0)$ ($x\equiv r/\xi_b$) 
and the rms radii are given 
for each of the constituent isotopes. 
Other parameters are fixed as $a_{bb}=4.29{\rm\,nm}$,  
$\omega_f =\omega_b=100{\rm\,Hz}$.}
\vspace{0.5cm}
\begin{tabular}{cccccccccc}
($N_{b},N_{f}$; $a_{bf}$)&atom&$\mu_{b}$\,or\,$\mu_{f}$&$E_{tot}/N$&$E_{kin}/N$&
    $E_{ho}/N$&$E_{BB}/N$&$E_{BF}/N$&$n (x=0)$&$\sqrt{\langle x^2 \rangle}$\\
    \hline
    
	($10^{3},10^{3};0.00$)&$^{39}$K&$4.19$&$3.66$&$1.16$&$1.96$&$0.53$&$0.00$&$98.7$&$1.40$\\
	                      &${}^{40}$K&$36.3$&$27.3$&$13.6$&$13.6$&$---$&$0.00$&$3.70$&$3.69$\\
	($10^{3},10^{3};2.51$)&$^{39}$K&$4.24$&$3.71$&$1.16$&$1.96$&$0.53$&$0.05$&$98.7$&$1.40$\\
	                      &${}^{40}$K&$36.4$&$27.3$&$13.6$&$13.7$&$---$&$0.05$&$3.47$&$3.70$\\
			      \hline
        ($10^{4},10^{3};0.00$)&$^{39}$K&$8.21$&$6.26$&$0.70$&$3.62$&$1.95$&$0.00$&$291$&$1.90$\\
                              &${}^{40}$K&$36.3$&$27.3$&$13.6$&$13.6$&$---$&$0.00$&$3.70$&$3.69$\\
	($10^{4},10^{3};2.51$)&$^{39}$K&$8.25$&$6.31$&$0.70$&$3.62$&$1.94$&$0.05$&$291$&$1.90$\\
                              &${}^{40}$K&$36.6$&$27.7$&$13.3$&$14.0$&$---$&$0.45$&$3.07$&$3.74$\\
    \end{tabular}
\end{table}
%
%
\begin{table}
\caption{
Results for the ${}^{41}$K-${}^{40}$K system. 
See Table I for the notation. 
The parameters are: 
$a_{bb}=15.13{\rm\,nm}$,  
$\omega_f =\omega_b =100{\rm\,Hz}$.}
\vspace{0.5cm}
\begin{tabular}{lccccccccc}
($N_{b},N_{f}$; $a_{bf}$)&atom&$\mu_{b}$\,or\,$\mu_{f}$&$E_{tot}/N$&$E_{kin}/N$&
    $E_{ho}/N$&$E_{BB}/N$&$E_{BF}/N$&$n (x=0)$&$\sqrt{\langle x^2 \rangle}$\\
    \hline
    	
	($10^{3},10^{3};0.00$)&$^{41}$K&$5.83$&$4.68$&$0.90$&$2.62$&$1.15$&$0.00$&$53.5$&$1.62$\\
	                      &${}^{40}$K&$36.3$&$27.3$&$13.6$&$13.6$&$---$&$0.00$&$3.70$&$3.69$\\
	
($10^{3},10^{3};-8.57$)&$^{41}$K&$5.65$&$4.49$&$0.91$&$2.62$&$1.16$&$-0.19$&$53.9$&$1.62$\\ 

&${}^{40}$K&$36.2$&$27.1$&$13.8$&$13.5$&$---$&$-0.19$&$4.13$&$3.68$\\ \hline
	($10^{4},10^{3};0.00$)&$^{41}$K&$13.0$&$9.58$&$0.50$&$5.65$&$3.43$&$0.00$&$135$&$2.38$\\
                              &${}^{40}$K&$36.3$&$27.3$&$13.6$&$13.6$&$---$&$0.00$&$3.70$&$3.69$\\		
	
($10^{4},10^{3};-8.57$)&$^{41}$K&$12.9$&$9.40$&$0.50$&$5.62$&$3.46$&$-0.18$&$136$&$2.37$\\                                                            

&${}^{40}$K&$35.4$&$25.6$&$14.7$&$12.7$&$---$&$-1.78$&$4.71$&$3.56$\\
    \end{tabular}
\end{table}
%
%
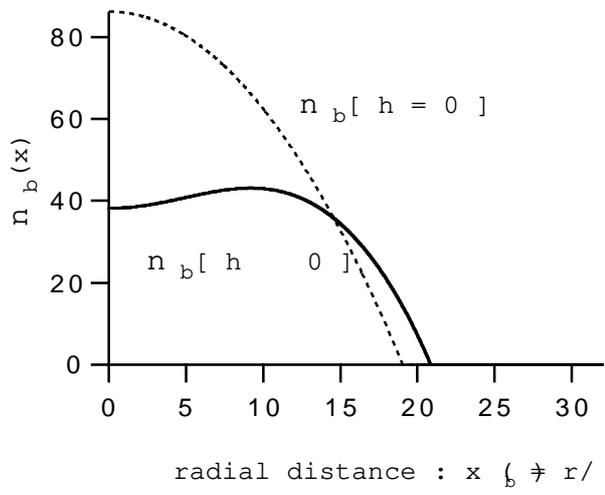
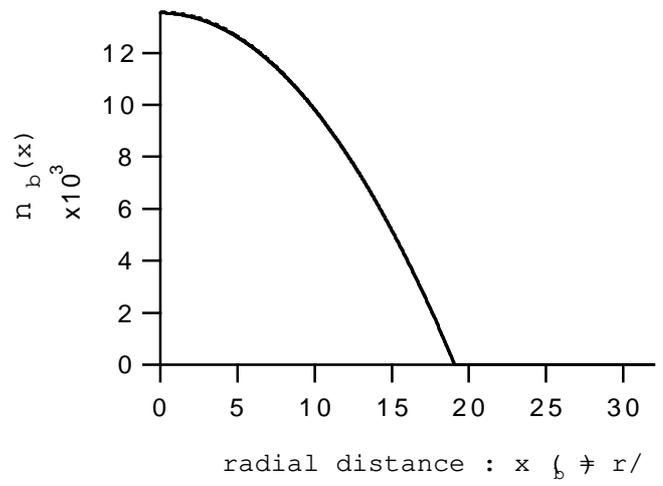
\begin{figure}
\caption{
Changes of the boson density distribution 
for different $N_f$ values: 
a) $N_f =10^6$ and b) $N_f =10^8$ 
with 
($\tilde{g} N_b$, $\tilde{h} N_f$, $\tilde{h} N_b$, 
$R_m$, $R_\omega$)
=
($4.22\times 10^{6}$, 
$4.22\times 10^{6} {\tilde{h}\over \tilde{g}}$,
$10^{6},4.22\times 10^{6} {\tilde{h}\over \tilde{g}}$, $1$, $1$). 
Solid lines are for $h/g =0.6$
and dotted lines are for $h/g =0$ (no boson-fermion interaction).}
\end{figure}

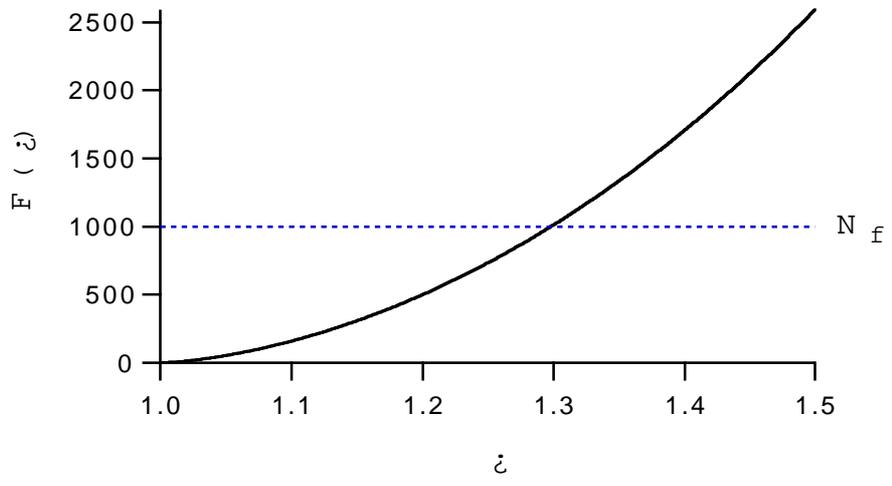
\begin{figure}
\caption{
Critical line of component separation for the system with
($\tilde{g} N_b$, $\tilde{h} N_f$, $N_f$, $\tilde{h} N_b$, 
$R_m$, $R_\omega$)
 =($3.0\times 10^4$, $3.0\times 10 h/g$, 
$10^3$, $3.0\times 10 h/g$, $1$, $1$).}
\end{figure}

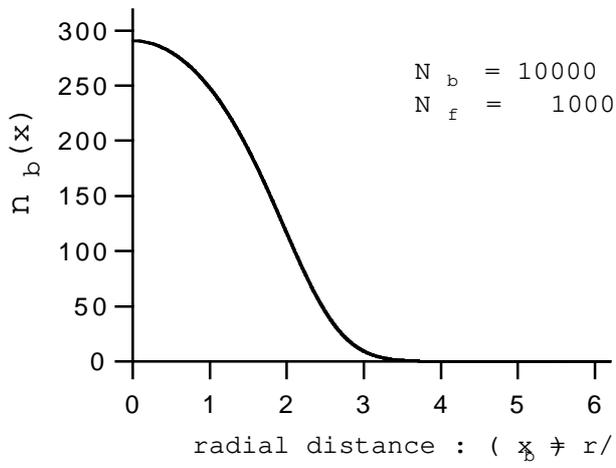
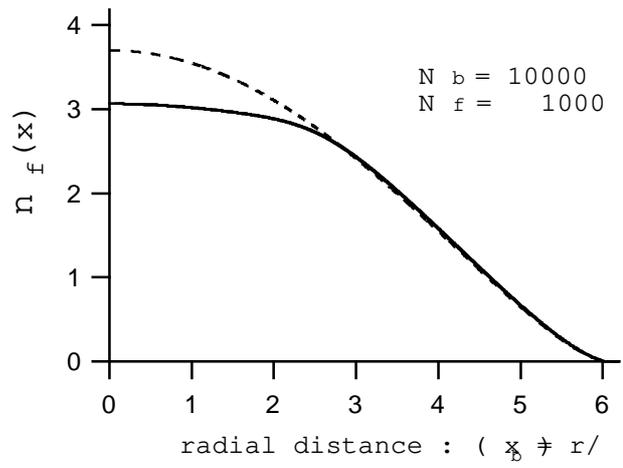
\begin{figure}
\caption{
Density distribution 
for the (${}^{39}$K, ${}^{40}$K) = (10000,1000) system: 
(a) for ${}^{39}$K and 
(b) for ${}^{40}$K (b) are plotted against 
$x \equiv r/\xi_b$. 
The solid line shows the full calculation 
with the (repulsive) boson-fermion interaction ($a_{bf}=2.51{\rm\,nm}$) 
and the dashed line without. 
Both lines are indistinguishable for the bosonic part (a).}
\end{figure}

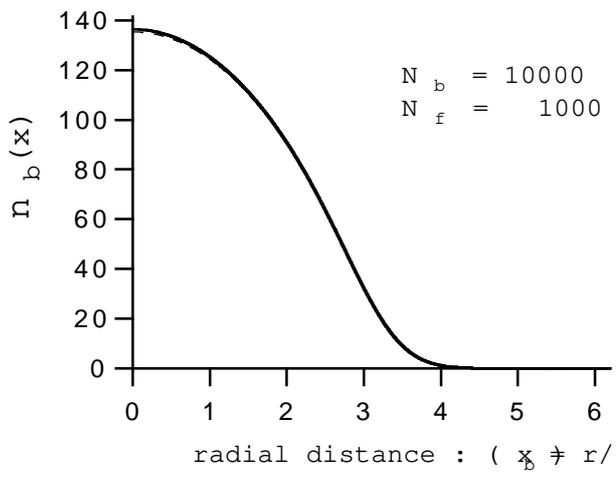
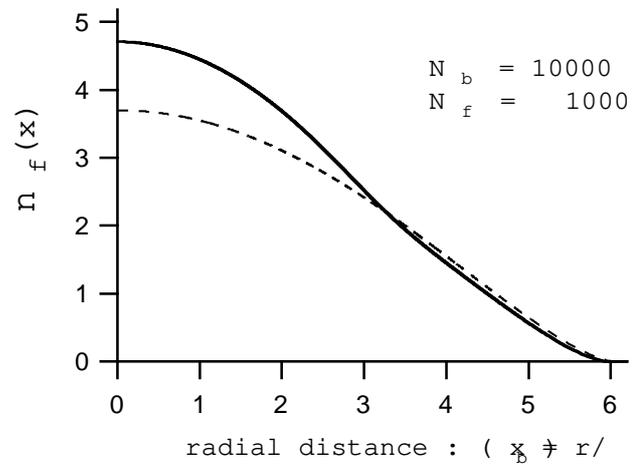
\begin{figure}
\caption{
Density distribution for the (${}^{41}$K,${}^{40}$K): 
(a) for the ${}^{41}$K and (b) for ${}^{40}$K. 
The boson-fermion interaction is attractive in this case 
($a_{bf} =-8.57{\rm\,nm}$).}
\end{figure}

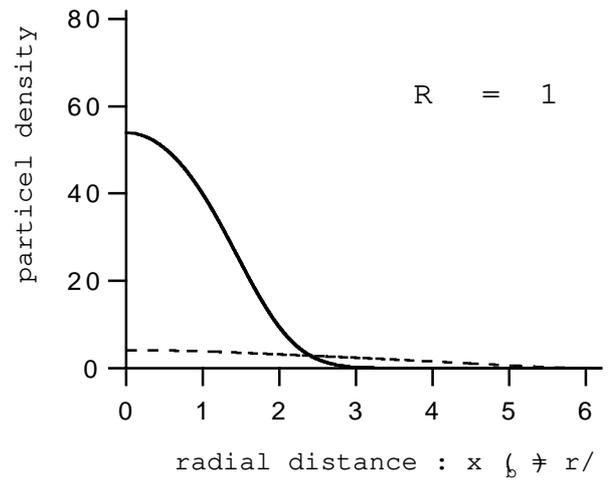
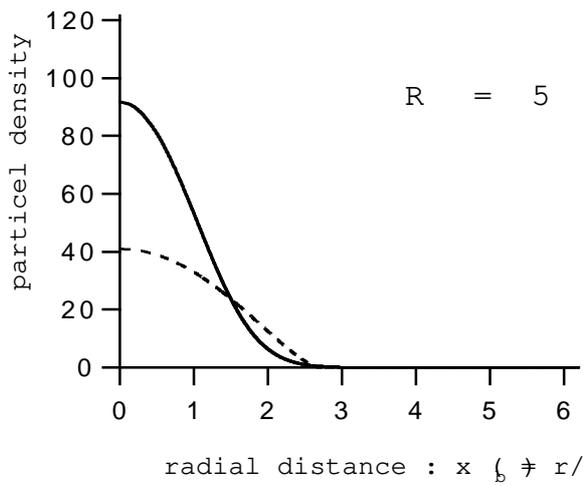
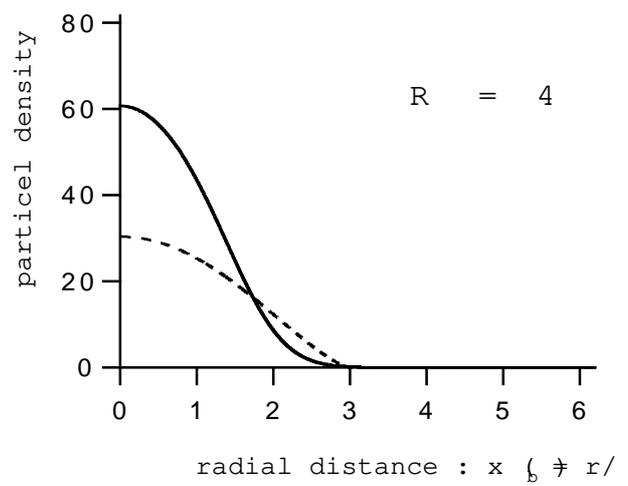
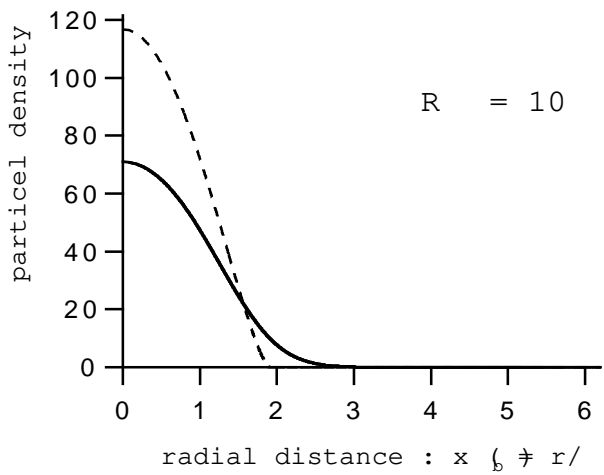
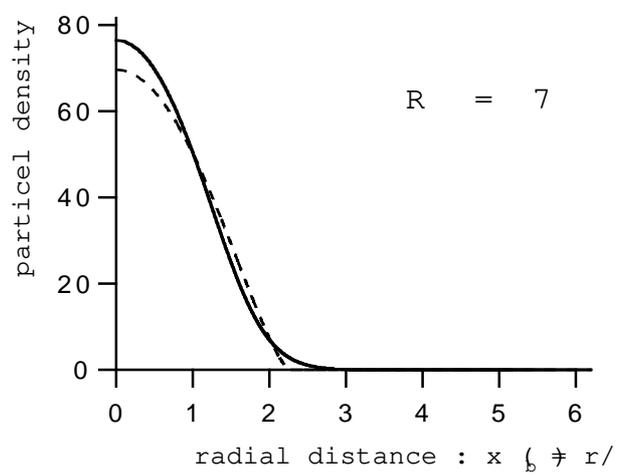
\begin{figure}
\caption{
Dependence of the density distribution 
on the oscillator-frequency ratio $R_\omega =\omega_f/\omega_b$: 
a) ${}^{39}$K-${}^{40}$K system with $R_\omega=1,5,10$ 
and b) ${}^{41}$K-${}^{40}$K system with $R_\omega=1,4,7$, 
where $(N_b,N_f)=(1000,1000)$ and $\omega_b =100{\rm\,Hz}$. 
The solid line shows the boson density $n_b(x)$ 
and the dashed line shows the fermion density $n_f(x)$. }
\label{Fig5}
\end{figure}
%

\begin{references}
\bibitem{FNA}
In \cite{ng}, $\alpha =0 \sim 2$ was used, 
where the phase separation was found in $\alpha =3/2 \sim 2$.
\bibitem{FNB}
The kinetic ($E_{kin}/N$), 
external potential ($E_{ho}/N$), 
boson-boson ($E_{BB}/N$) and 
boson-fermion ($E_{BF}/N$) interaction energies per particle 
are $1.08$, $2.15$, $1.59$, and $-1.84$, respectively, 
in th unit of $\hbar \omega_b /2$.
\end{references}
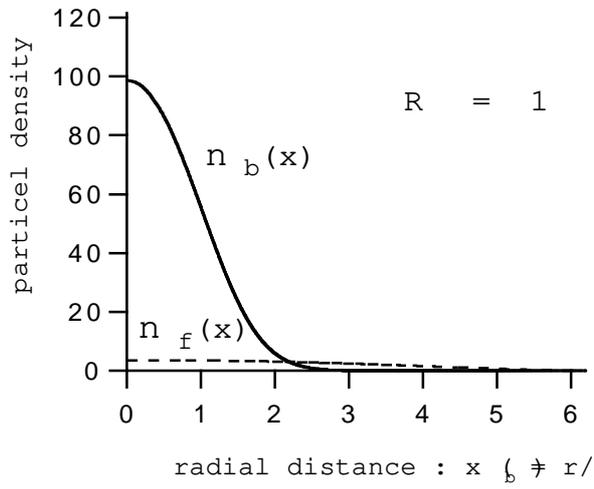
\end{document}